\documentclass[fleqn,usenatbib]{mnras}

\usepackage{newtxtext,newtxmath}

\DeclareRobustCommand{\VAN}[3]{#2}
\let\VANthebibliography\thebibliography
\def\thebibliography{\DeclareRobustCommand{\VAN}[3]{##3}\VANthebibliography}

\usepackage{fix-cm}
\usepackage{bm}

\usepackage{graphicx}	

\title[Star formation in Eos]{Searching for star formation towards the Eos molecular cloud}

\author[Saxena et al.]{
Suryansh Saxena,$^{1}$\thanks{E-mail: s.saxena@se24.qmul.ac.uk}
Thomas J. Haworth,$^{1}$\thanks{E-mail: t.haworth@qmul.ac.uk},
Blakesley Burkhart$^{2,3}$,
Thavisha Dharmawardena$^{3}$,
\newauthor
Edward Gillen$^{1}$,
Kate Pattle$^{4}$,
Janik Karoly$^{4}$,
Erika Hamden $^{5}$
\\
$^{1}$Astronomy Unit, Department of Physics and Astronomy, Queen Mary University of London, Mile End Road, London, E1 4NS, UK\\
$^{2}$Department of Physics and Astronomy, Rutgers University,  136 Frelinghuysen Rd, Piscataway, NJ 08854, USA\\
$^{3}$Center for Computational Astrophysics, Flatiron Institute, 162 Fifth Avenue, New York, NY 10010, USA \\
$^{4}$Department of Physics and Astronomy, UCL, Gower St, London WC1E
6BT, UK \\
$^{5}$Steward Observatory, University of Arizona, 933 N Cherry Ave, Tucson, AZ 85719, USA \\
}

\date{Accepted XXX. Received YYY; in original form ZZZ}

\pubyear{\the\year{}}

\begin{document}
\label{firstpage}
\pagerange{\pageref{firstpage}--\pageref{lastpage}}
\maketitle

\begin{abstract}
The Eos cloud, recently discovered in the far ultraviolet  via H$_2$ fluorescence, is one of the nearest known dark molecular clouds to the Sun, with a distance spanning from $\sim94-136$\,pc. However, with a mass ($\sim5.5\times10^3$\,M$_\odot$) just under $40\,$per cent that of star forming clouds like Taurus and evidence for net molecular dissociation, its evolutionary and star forming status is uncertain.  We use Gaia data to investigate whether there is evidence for a young stellar population that may have formed from the Eos cloud. Comparing isochrones and pre-main sequence evolutionary models there is no clear young stellar population in the region. While there are a small number of $<10$\,Myr stars, that population is statistically indistinguishable from those in similar search volumes at other Galactic latitudes. We also find no unusual spatial or kinematic clustering toward the Eos cloud over distances $70-150$\,pc. Overall we conclude that the Eos cloud has most likely not undergone any recent substantial star formation, and further study of the dynamics of the cloud is required to determine whether it will do so in the future. 
\end{abstract}




\section{Introduction}\label{sec1}

Nearby star forming regions such as Taurus and Lupus have played an important role in developing our understanding of star and planet formation \citep[e.g.][]{1978ApJ...224..857E, 1995ApJS..101..117K, 2008hsf2.book..295C, 2013A&A...550A..38P, 2015ApJ...808L...3A, 2016ApJ...828...46A}. Thanks to Gaia \citep{2016A&A...595A...1G}, recent years have seen a huge development in our understanding of the 3D structure of the nearby Galactic interstellar medium and star forming regions \citep[e.g.][]{2022Natur.601..334Z, 2024ApJ...973..136O, 2024MNRAS.532.3480D, 2025arXiv250400093Z}. Particularly significant was the remarkable discovery that the Sun sits towards the centre of a supernova driven bubble that is likely associated with our nearest star forming regions \citep{2022Natur.601..334Z}. However, our view of the local bubble has recently been updated again with the discovery of the ``Eos'' cloud \citep{EosCloudPaper}. At a distance of just 94\,pc Eos is one of the closest known molecular clouds to the Sun, but much of it was previously undiscovered because it is over 99\% CO-dark. \cite{EosCloudPaper} first detected the Eos cloud using data from FIMS/SPEAR  \citep{2011ApJS..196...15S} via H$_2$ fluorescence. 

\cite{EosCloudPaper} calculated a total cloud mass for Eos of $\sim5.5\times10^3$\,M$_\odot$ (H$_2$ mass $\sim3.4\times10^3$\,M$_\odot$), just under $40\,$per cent that of the total mass of the Taurus/Lupus clouds, and estimated that it is globally marginally Jeans stable. Using the methodology outlined in \cite{2025ApJ...982...24B}, they also computed global H$_2$ formation and dissociation rates for the cloud, finding it is likely net dissociating. However, a global analysis of the cloud is simplified and the cloud is certainly non-uniform, with a small CO-bright condensation referred to as MBM 40 \citep{1996ApJ...465..825M, 2003NewA....8..795C, 2013MNRAS.436.1152C, 2022A&A...668L...9M, 2023A&A...676A.138M} located at the more distant lower latitude portion of the cloud. The Jean's analysis is therefore insufficient to rule out previous or future star formation. \cite{EosCloudPaper} did not study stellar populations associated with the cloud and it is important to determine whether the Eos cloud is a site of recent or ongoing star formation. 

Gaia has revolutionized our ability to study stellar populations and identify clusters and sub-clustered groupings through spatial and kinematic association \citep[e.g.][]{2019A&A...630A.137G, 2024MNRAS.534.2566S}. In this letter, our goal is to use data from Gaia to search for signs of ongoing or recent (e.g. within the last $20$\,Myr) star formation towards the Eos cloud.

\section{Data and Methodology} \label{sec1}

\subsection{Gaia sample and refinement}
We use Gaia DR3 data to study the stellar population towards the Eos cloud \citep{2016A&A...595A...1G, 2021A&A...649A...1G}. The DR3 data release has improved photometric data compared to previous releases, a much expanded radial velocity survey, and a very extensive astrophysical characterisation of Gaia sources \citep{2023A&A...674A...1G}. 

The Eos cloud extends around 20 degrees in latitude and longitude. To account for the possibility that stars formed within the Eos cloud but migrated out of it, we extracted all stars within a 25 degree circular search radius centered on the Eos cloud, at (RA, DEC) = (15h46m24.58s, +21d42m44.72s), which corresponds to Galactic (l,b) = (35, 50) degrees \citep{EosCloudPaper}. The Eos cloud starts at a distance of $\sim94\,$pc and extends out to $\sim136\,$pc. We therefore sample stars in the range $70-150$\,pc which corresponds to Gaia parallaxes in the range 6.666$-$14.285\,mas.  This results in a sample of 41436 stars. 

\begin{figure}
    \centering
    \includegraphics[width=0.9\linewidth]{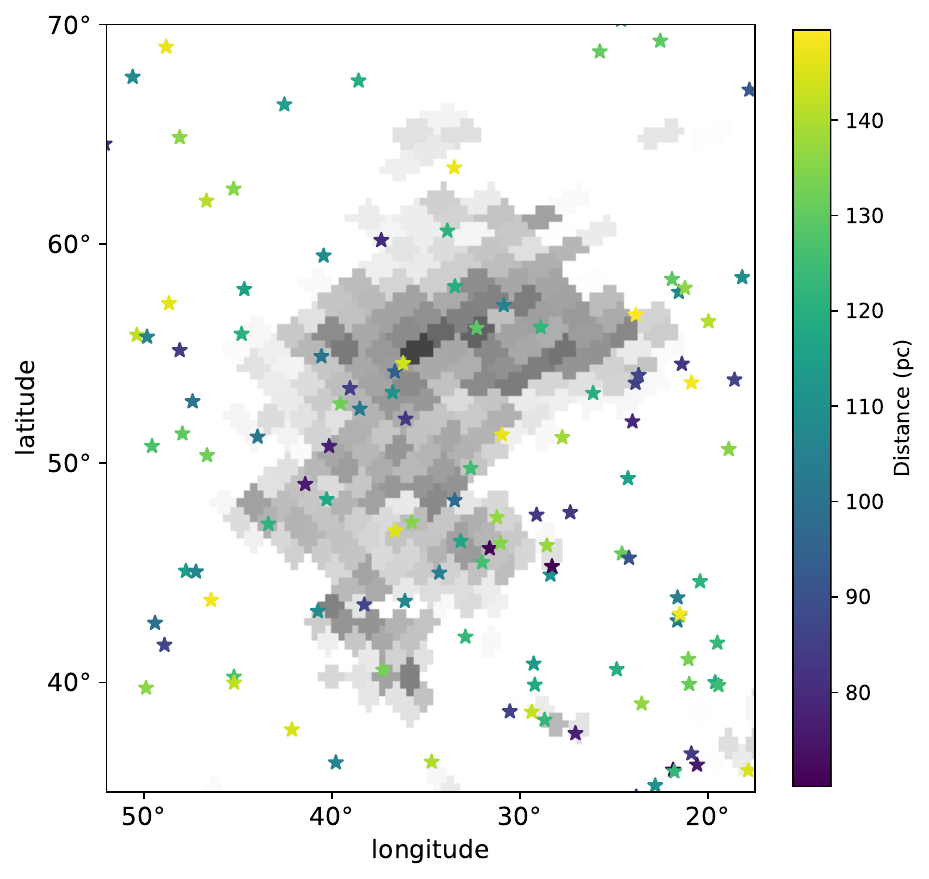}
    \caption{Distribution of stars less than 20 Myr age in the Eos region. The star markers are the dataset pulled from Gaia DR3 colored by distance, which are overplotted upon a FIMS/SPEAR grayscale map of the ratio of H$_2$ intensity to the total FUV intensity which shows the on-sky extent of the Eos cloud.  }
    \label{spatial distribution}
\end{figure}

For illustrative purposes, the stars in the Gaia sample that are $<20$\,Myr (discussed in \ref{sec:PMSevo})  are plotted in Figure \ref{spatial distribution}, where the grayscale is the FIMS/SPEAR  ratio of H$_2$ intensity to total far-ultraviolet (FUV) intensity \citep{EosCloudPaper} and the colourscale is the Gaia parallax distance to the stars. From Figure \ref{spatial distribution} there is no clear evidence for spatial clustering in 2D projection, nor based on the 3D coordinates derived from the parallaxes. Furthermore, the Eos cloud is at larger distances with increasing latitude, and this is not reflected in the stellar distances. We therefore focus our subsequent discussion on ages, kinematics and comparison with other stellar populations. We also note that none of the stars in Figure \ref{spatial distribution} are coincident with the CO-bright region discussed in \citep{EosCloudPaper}.

Since Eos is situated nearby and at high latitude, extinction and reddening effects  in the Gaia passbands\footnote{\url{https://www.cosmos.esa.int/web/gaia/edr3-passbands}} are small \citep{paunzen2022detection}. Nevertheless, we corrected the absolute magnitude $M_\textrm{G}$ and the $(G_{\textrm{BP}}-G_{\textrm{RP}})$ colour for extinction and reddening \citep{kordopatis2023stellar}. For completeness, we also accounted for parallax error following \cite{groenewegen2021parallax} and computed distances from parallaxes following \citep{Bailer-Jones2015} which, given the targets are nearby, resulted in only a modest adjustment to the parallax values and corresponding distances, respectively.

\subsection{Pre-main sequence evolutionary models}
\label{sec:PMSevo}

Our goal is to identify any young stellar population that may be associated with star formation in the Eos cloud. For interpreting the Gaia photometry we adopt the pre–main sequence evolutionary models from \cite{baraffe2015new}, which are non-magnetic models. Other non-magnetic PMS models include the Pisa models \citep{tognelli2011pisa}, MIST models \citep{choi2016mesa}, and Dartmouth models \citep{dotter2008dartmouth}. The effect of magnetic fields have also been included, e.g. magnetic inhibition of convection \citep{feiden2016}, starspots \citep{Somers2020}, and the empirical approach of \cite{stassun2014empirical}. The magnetic models typically predict  older ages for similar stellar parameters \citep[e.g.,][]{feiden2016magnetic}, and thus our choice of the \cite{baraffe2015new} model is conservative in the context of searching for a young stellar population, in that if we do not find a young population magnetic models would only conclude the same population is slightly older.

\section{Results} \label{sec:result}
\subsection{Color Magnitude Diagram}
\begin{figure*}
    \centering
    \includegraphics[width=0.7\linewidth]{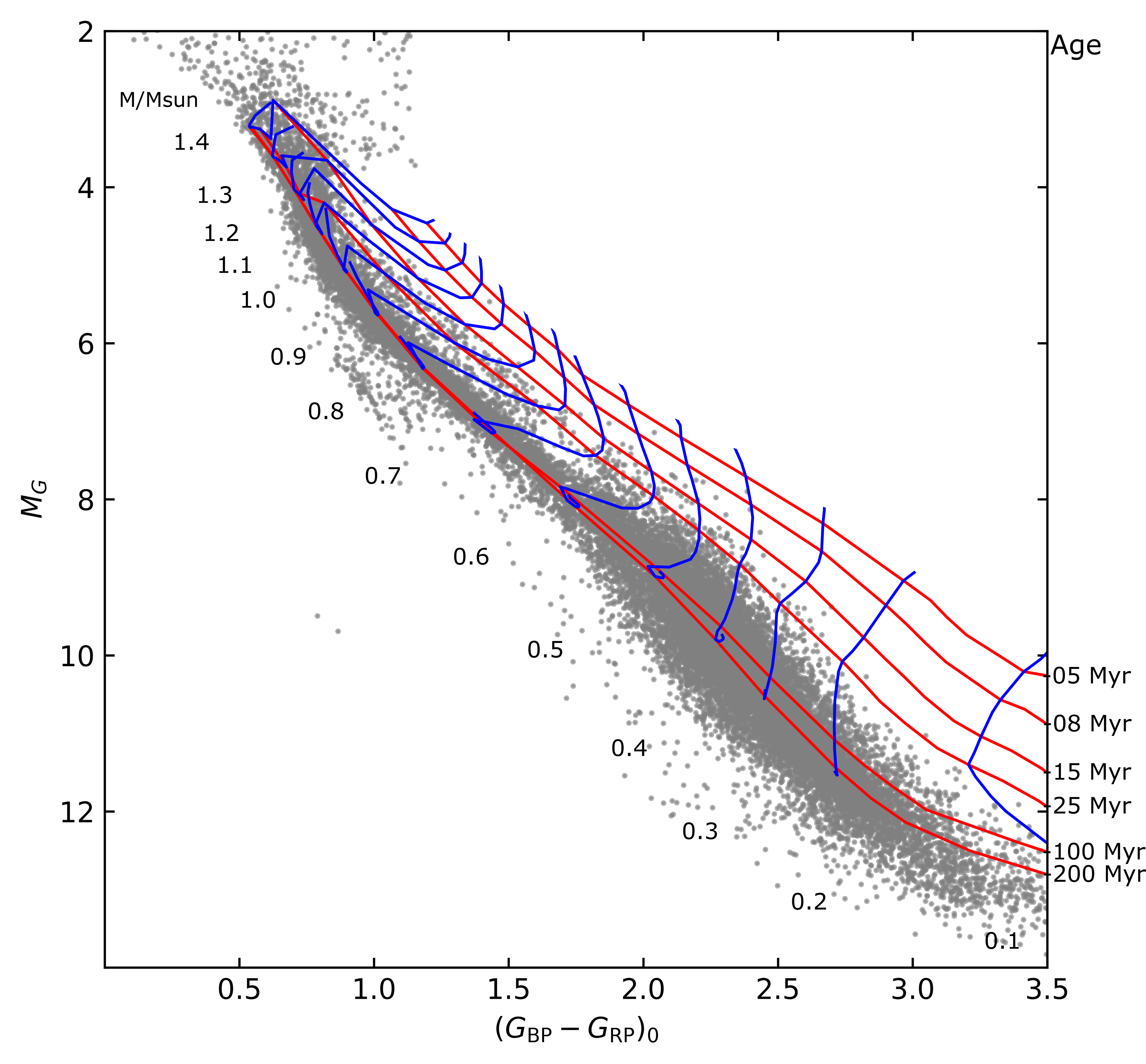}    
    \vspace{-0.3cm}
    \caption{Color magnitude plot with isochrones (red) and pre-main sequence stellar tracks (blue) from \protect\cite{baraffe2015new}. Here, the gray points represents the stellar population in a 25 degree search radius over distances 70-150\,pc towards the Eos cloud. }
    \label{multiage}
\end{figure*}

To analyze the age distribution of stars in the Eos Cloud, we constructed a Color-Magnitude Diagram (CMD) using Gaia DR3 photometric data. The CMD plots the absolute G magnitude ($M_\textrm{G}$) against the ($G_{\textrm{BP}}$-$G_{\textrm{RP}}$)$_0$ color index, providing a clear visualization of the stellar population in the region. 
Figure \ref{multiage} shows a CMD for our extinction, reddening and parallax-corrected stellar sample. Overlaid are \cite{baraffe2015new} PMS evolutionary tracks and isochrones. Other star forming regions on the edge of the local supernova-driven bubble have typical stellar ages $<3$\,Myr \citep[e.g. for Taurus, Lupus, Ophiuchus, Chameleon;][]{2008hsf2.book..295C, 2021A&A...646A..46G, 2022A&A...667A.163M, 2023AJ....165...37L}. Conversely, towards the Eos cloud there are no stars at all in that age range over the $70-150$\,pc distance we considered (Eos spans 94-136\,pc). There is a population of $<20$\,Myr stars, however we will demonstrate in section \ref{adjacent} that a population similar to this is found in other similar search volumes at this galactic latitude. 
Furthermore, it is worth noting that the model isochrones are for single stars and that, at a given age, equal-mass binary systems will form a sequence parallel to the single star sequence but $\sim$0.75 mag brighter.
The Gaia CMD hence demonstrates convincingly that star formation is not currently actively taking place in the Eos cloud. 

\subsection{Proper Motion Analysis}

\begin{figure}
    \centering
    \includegraphics[width=0.9\linewidth]{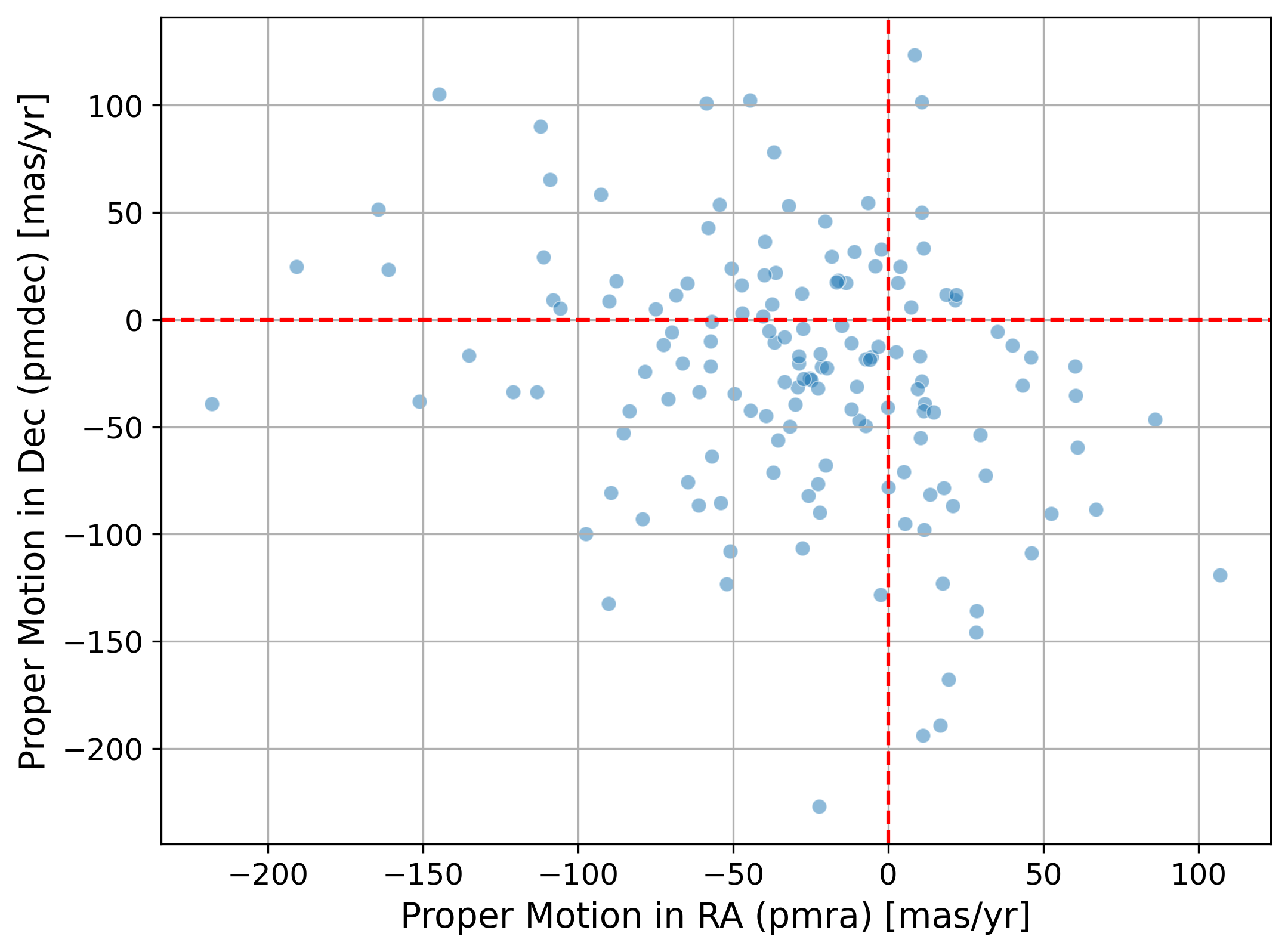}
    \vspace{-0.3cm}
    \caption{Proper motions of the sources towards the Eos cloud that are $\leq20$\,Myr.  The red lines indicates a zero reference point relative to the solar position.}
    \label{propermotion}
\end{figure}

Despite our finding no evidence for spatial clustering or a very young stellar population, an older coherent population might be identified using the stellar kinematics \citep[e.g.][]{2024MNRAS.534.2566S}. By analyzing the velocity dispersion of stars in the Gaia sample, we can determine whether the Eos cloud exhibits characteristics of a bound stellar system, a young star-forming region, or a dispersed field of unassociated stars.

Figure \ref{propermotion} shows the proper motion in right ascension ($\mu_{\alpha *}$) and declination ($\mu_\delta$) for all stars of $<20$\,Myr age. There is a broad and highly dispersed motion pattern and we confirm a lack of clustering with a k-mean test. Unlike young stellar clusters, which typically display a narrow and coherent velocity distribution due to strong gravitational binding \citep{baumgardt2019mean}, the stars toward the Eos cloud exhibit a significantly broad range of proper motions. This suggests that the stellar population is not gravitationally bound and is more characteristic of an older, dynamically evolved system \citep{bland2016galaxy}. 

\begin{figure}
    \centering
    \includegraphics[width=0.9\linewidth]{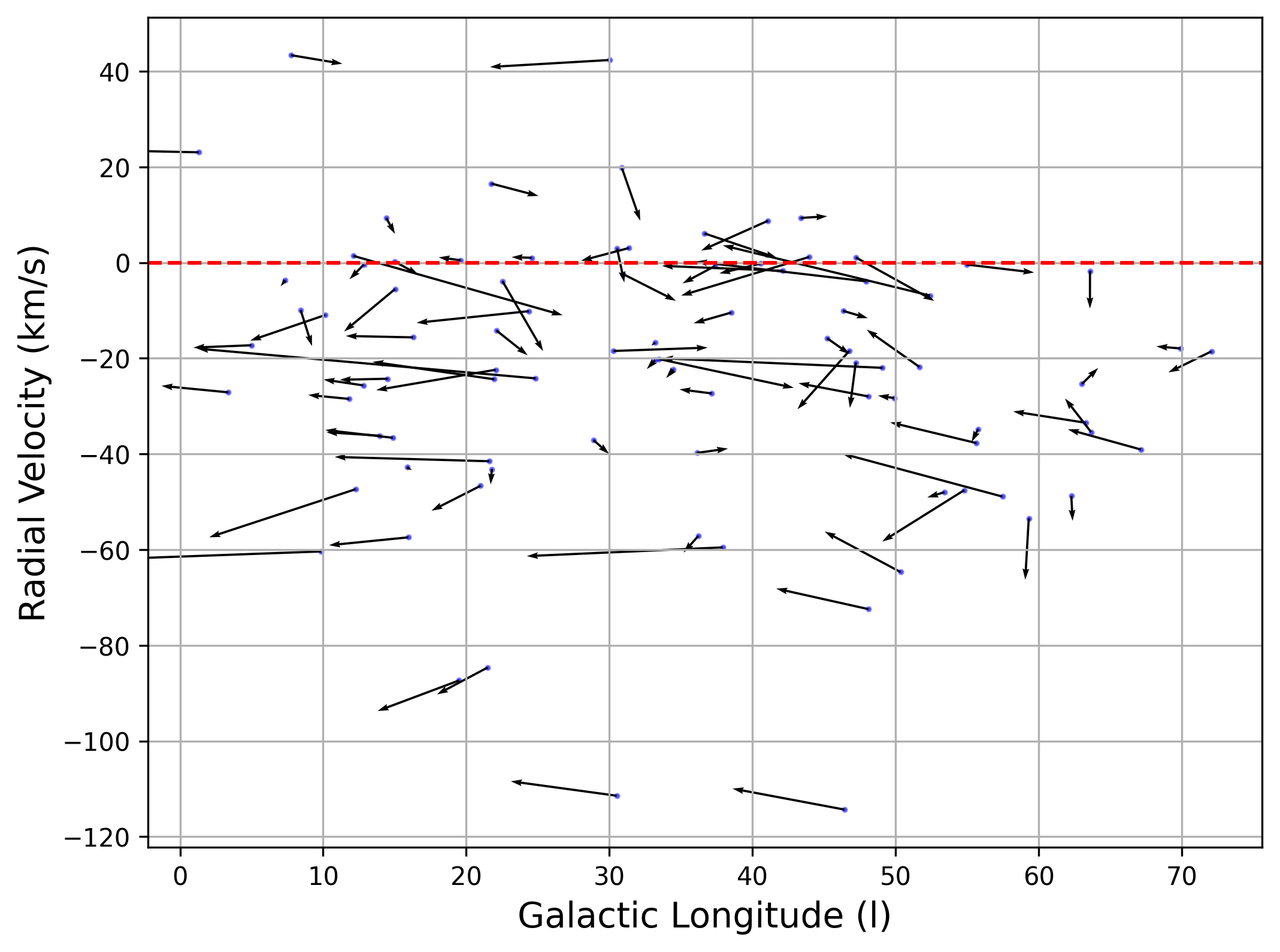}
    \caption{Line of sight radial velocity distribution for sources towards the Eos cloud that are $\leq20$\,Myr. }
    \label{los}
\end{figure}

The line of sight (LOS) radial velocity distribution for the Eos cloud is shown in Figure \ref{los}. The scatter in LOS provides further evidence of no clustering in the Eos cloud. 

\begin{figure*}
        \includegraphics[width=0.45\linewidth]{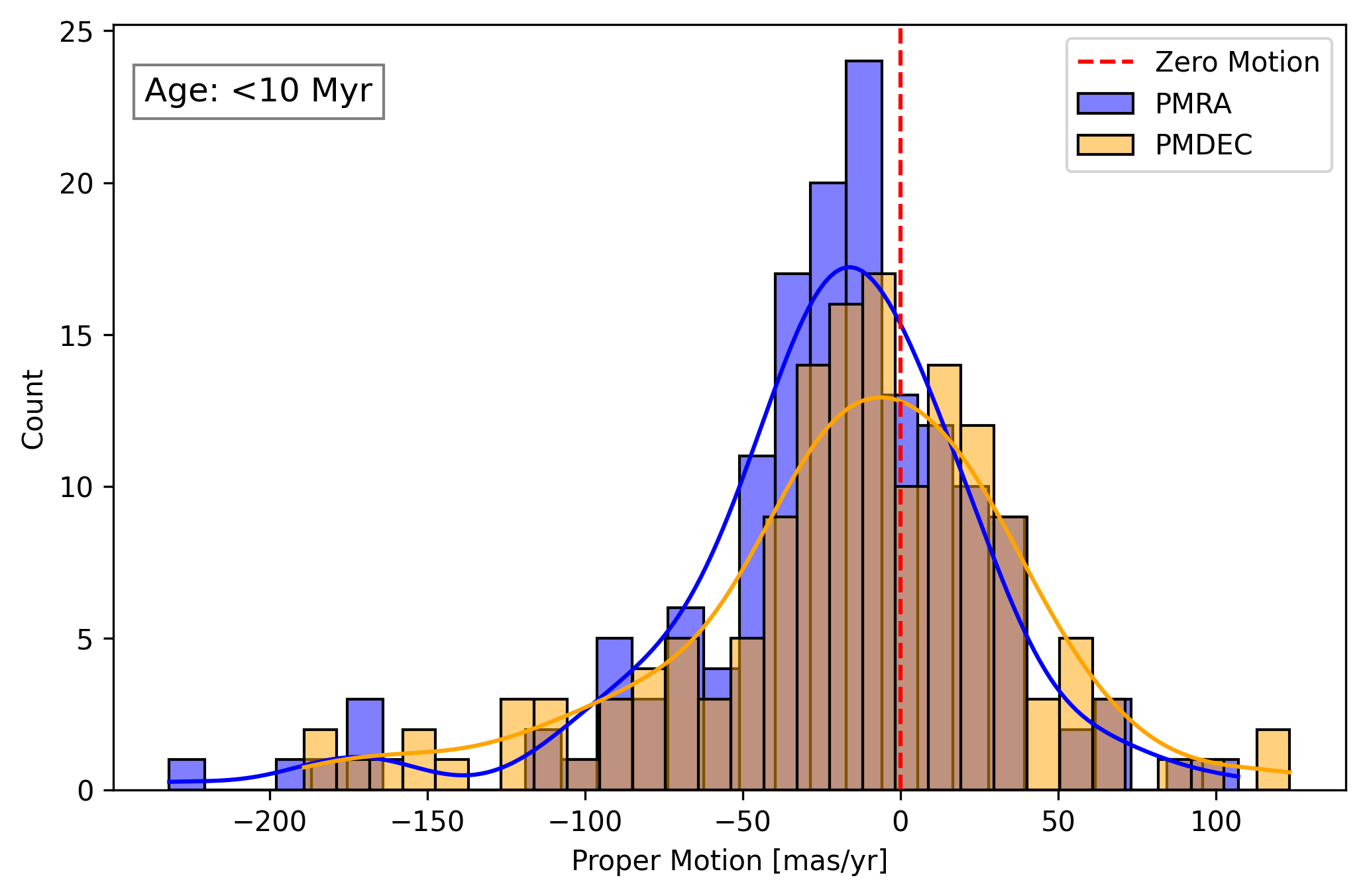}
        \includegraphics[width=0.45\linewidth]{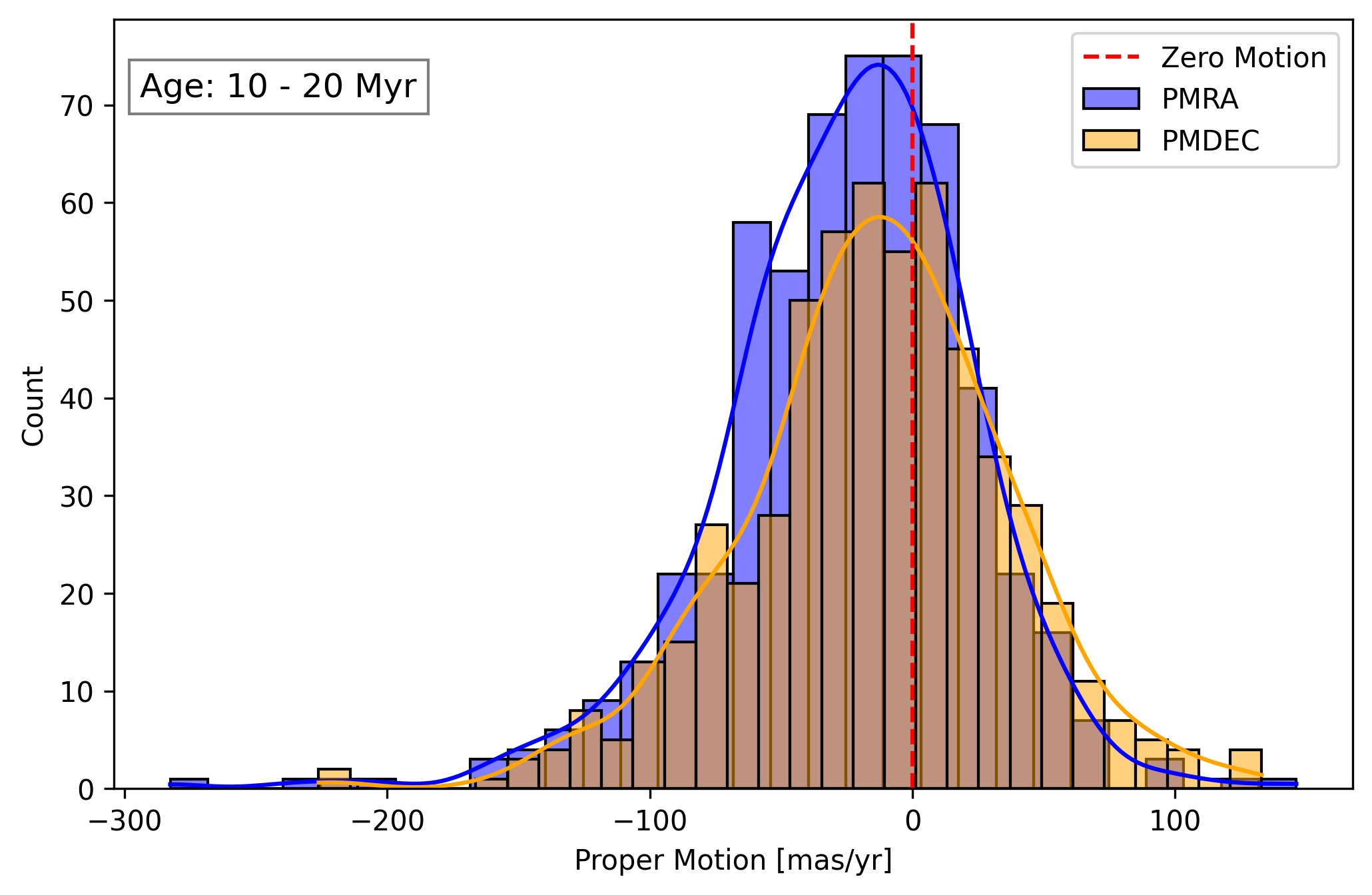}
        \includegraphics[width=0.45\linewidth]{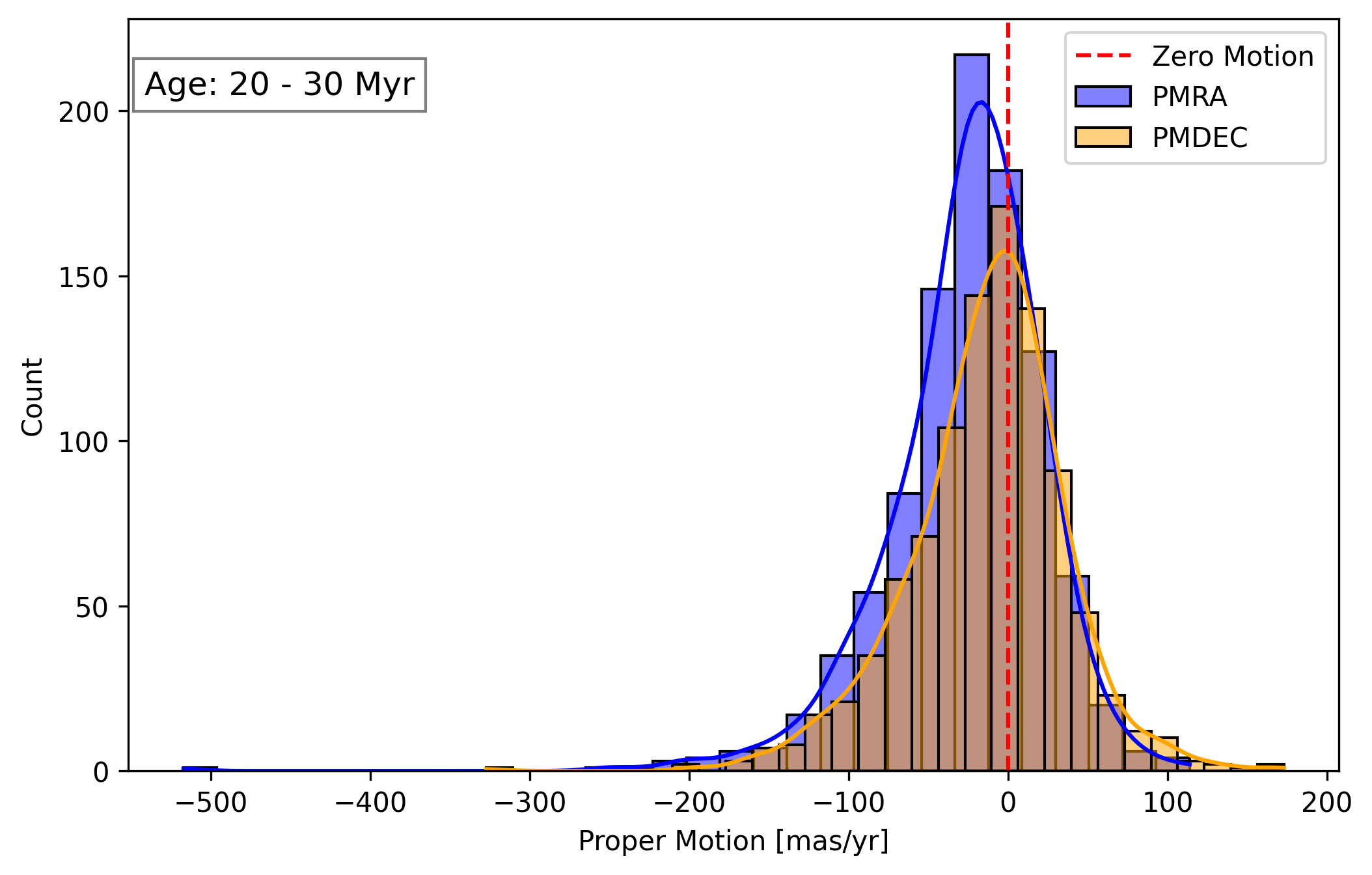}
        \includegraphics[width=0.45\linewidth]{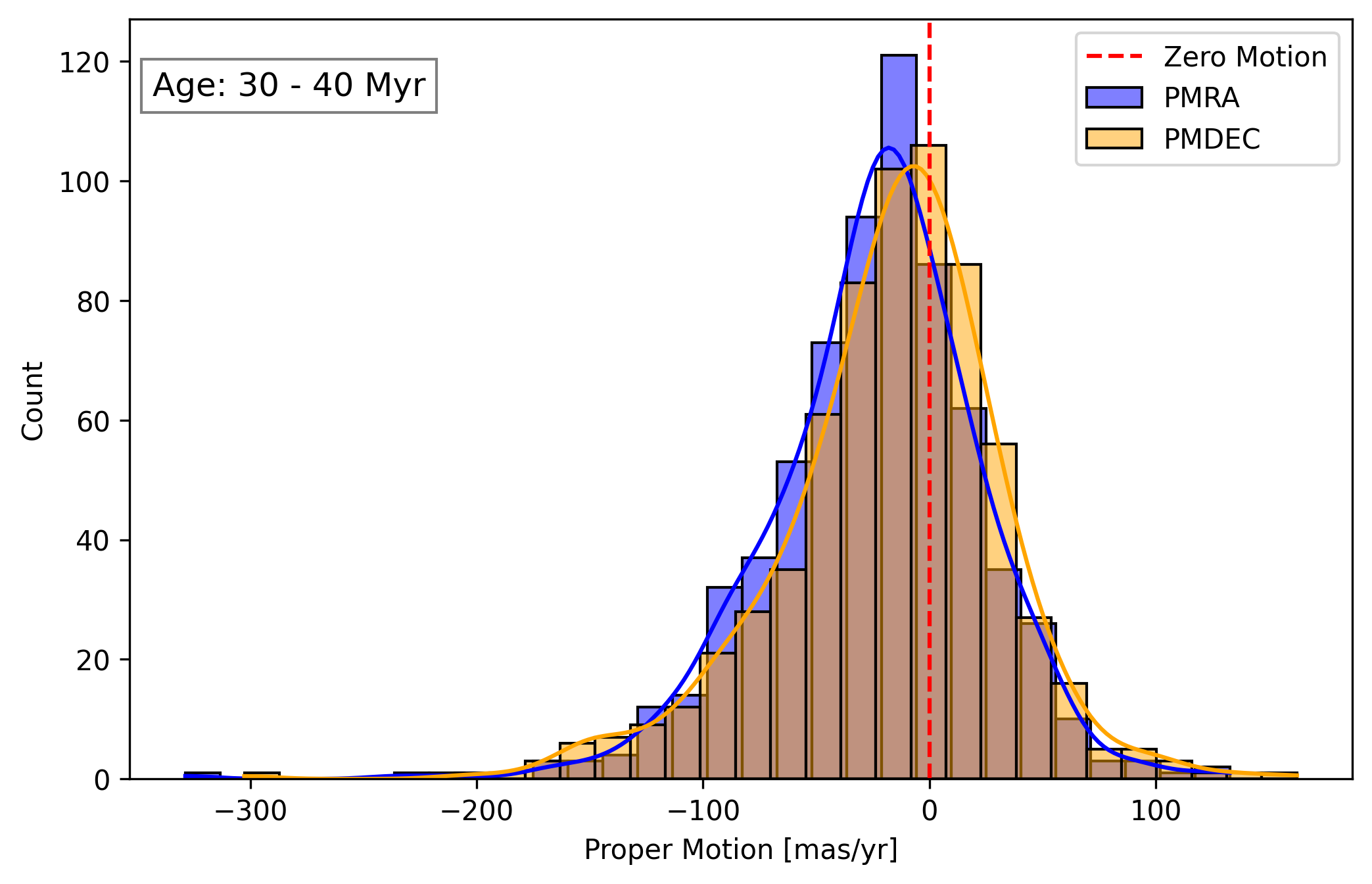}
        \includegraphics[width=0.45\linewidth]{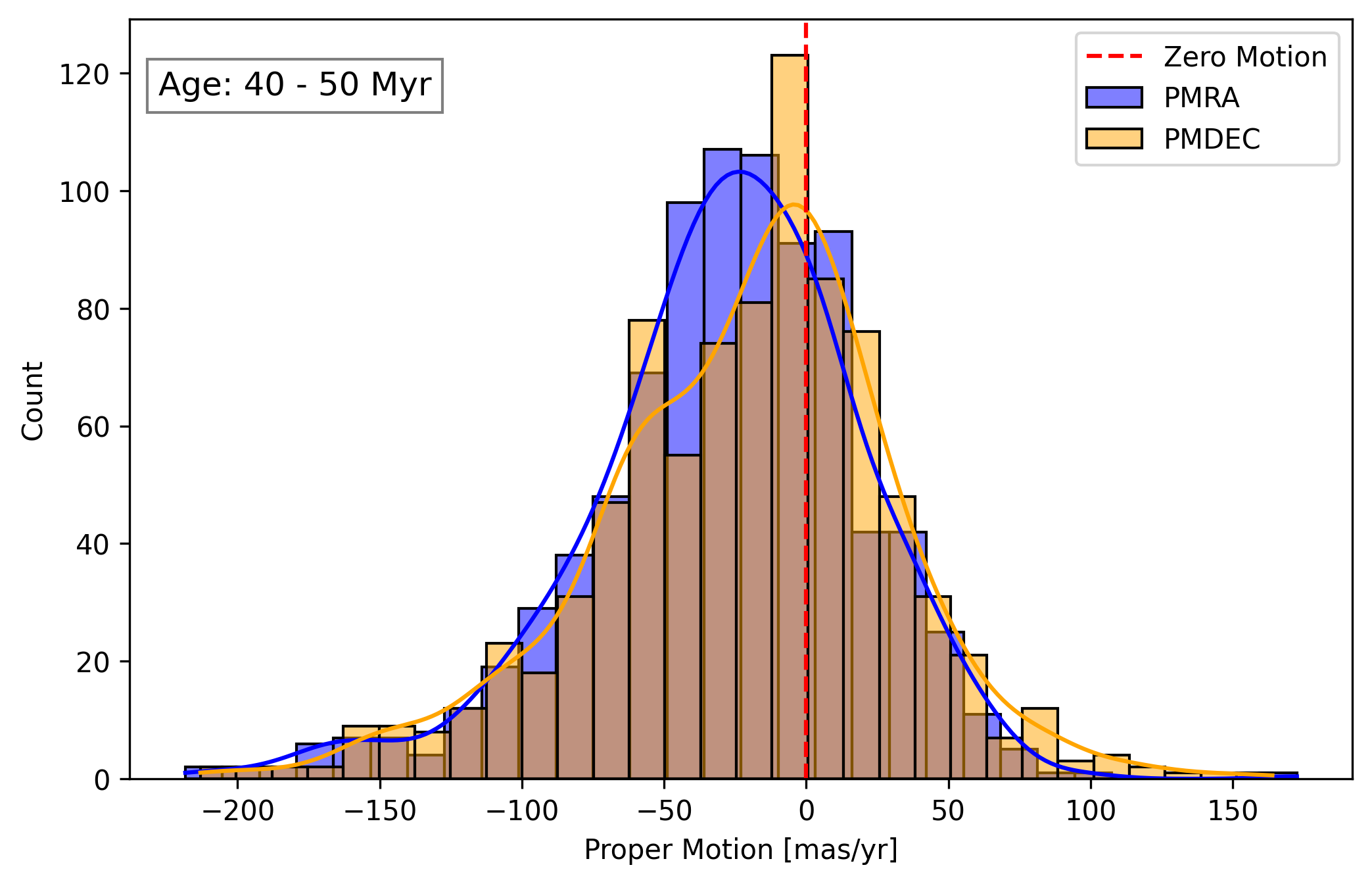}
        \includegraphics[width=0.45\linewidth]{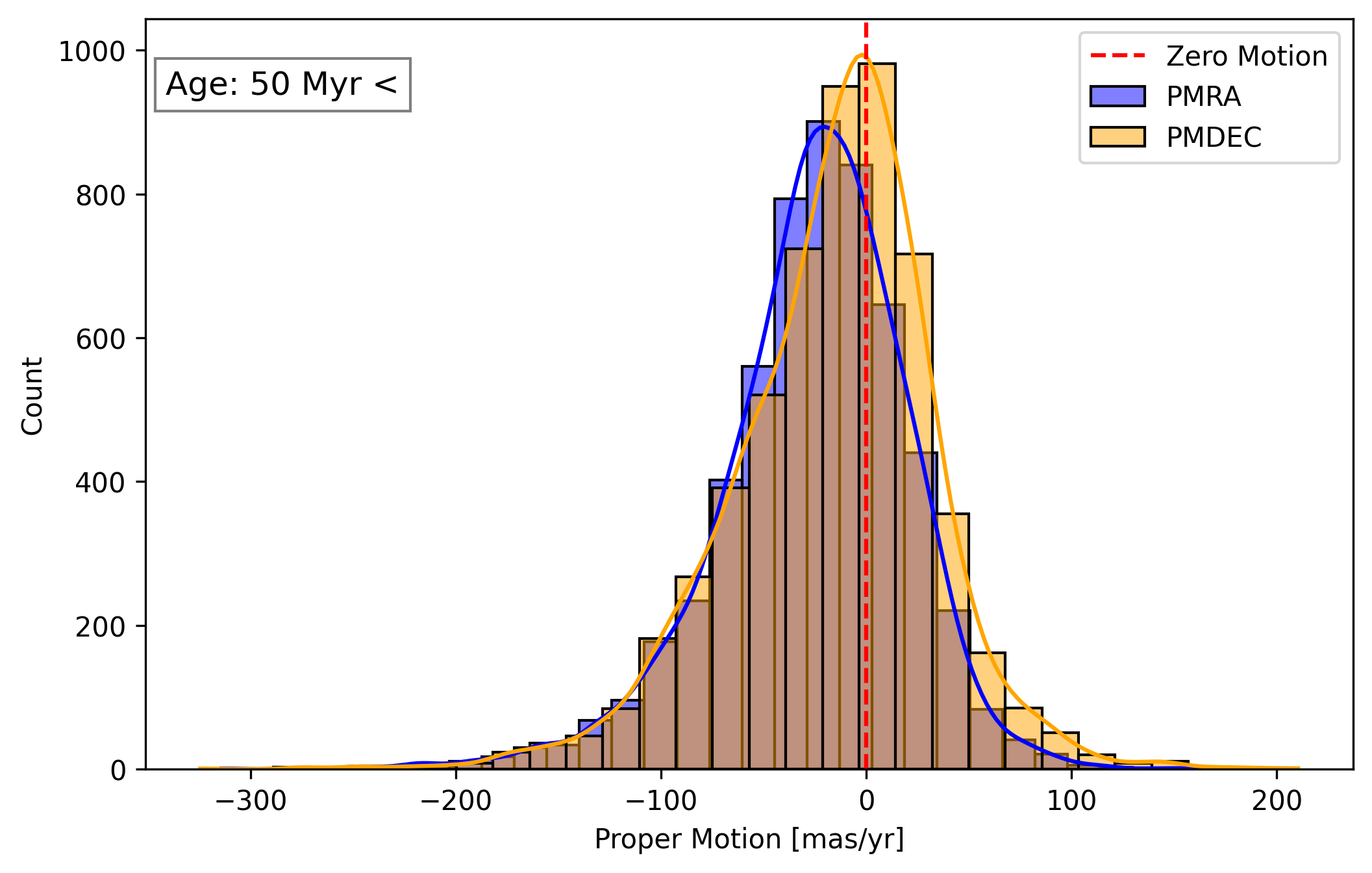}
    \vspace{-0.2cm}
    \caption{Proper motion histogram plotted using Gaia sample separated based on age. Red line indicates a zero reference point relative to the solar position.}
    \label{multihist}
\end{figure*}

To further explore the internal motion trends, we separated the entire stellar population based on age. Figure \ref{multihist} shows histograms of proper motions for different stellar age groups.  The typical spread in proper motions in both $\mu_{\alpha*}$ and $\mu_\delta$ is -100 to +100 mas/yr, confirming that there is no dominant motion direction. This dispersion pattern is inconsistent with young, gravitationally bound clusters. Instead, the broad distribution is indicative of a population that has been subject to significant dynamical evolution, possibly due to Galactic tidal forces or past interactions with other stellar structures \citep{kroupa2001variation, bovy2017stellar}. 

\subsection{Adjacent Regions}\label{adjacent}
As a final check we compared the stellar population towards the Eos cloud with analogous search volumes at the same Galactic latitude, but offset in Galactic longitude by 25 degrees to the left (negative longitude) and right  (positive longitude) from the edges of the Eos search volume described above. A total of 40848 objects were observed in the left dataset and 38579 objects in the right dataset.

\begin{figure*}
    \centering
    \includegraphics[width=0.8\linewidth]{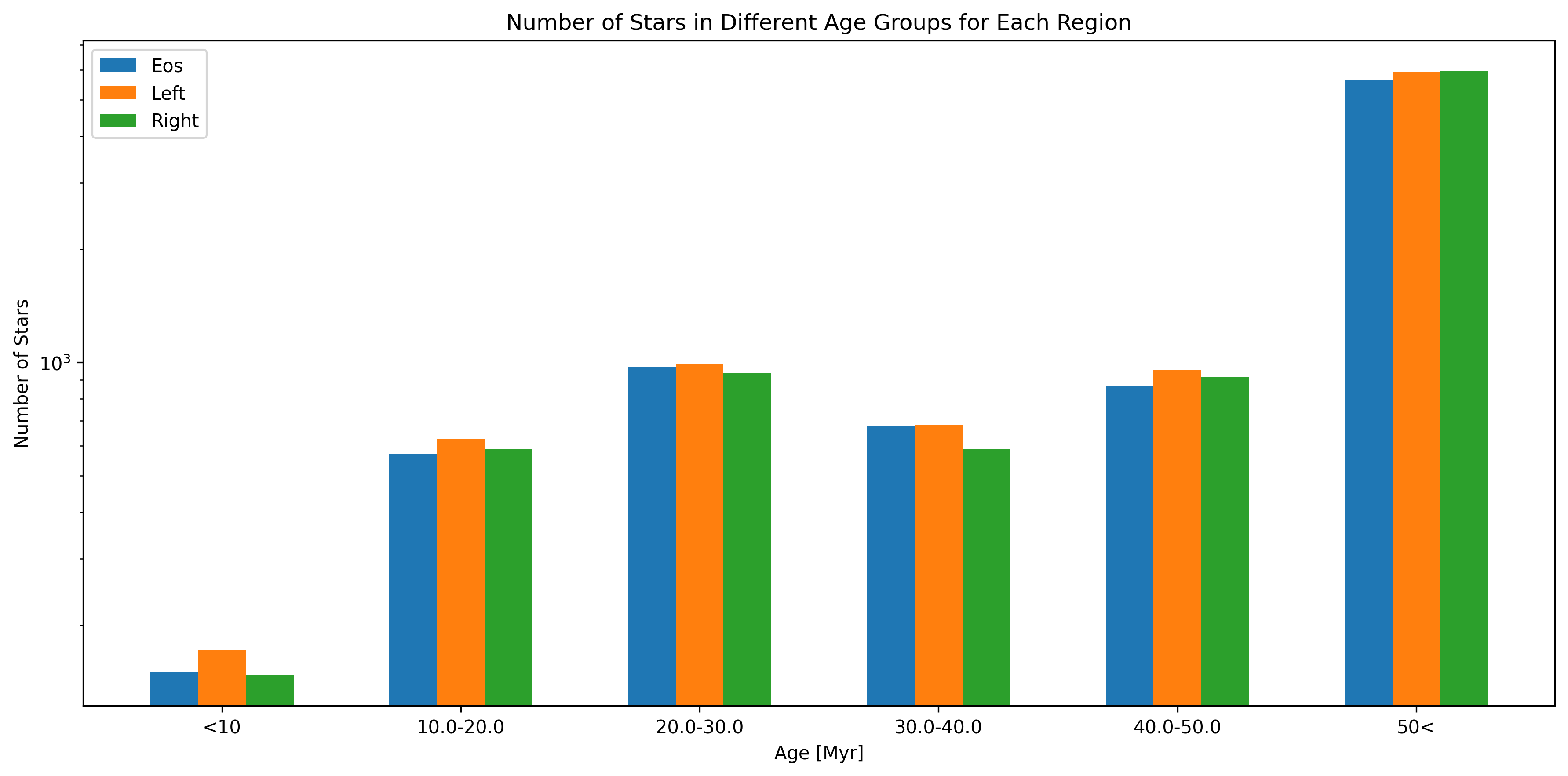}
    \vspace{-0.2cm}
    \caption{Histogram comparing the number of stars by age group for three 25 degree radius search fields. One is centered on the Eos cloud and the others are to the left and right of the Eos search field (negative and positive longitudes, respectively). We find no evidence for the stars in the field towards the Eos cloud being distinct from those in adjacent fields. }
    \label{agecomparison}
\end{figure*}

A histogram comparing the number of stars of different ages  towards the Eos cloud and adjacent regions is shown in Figure \ref{agecomparison}. They are all similar and the population towards the Eos cloud shows no evidence for an excess of young stars. The young population is consistent with other similar volumes. The spatial distributions, proper motion histograms, and age distributions all exhibit similar trends across the three regions, suggesting that the cloud does not represent a distinct or isolated stellar grouping. Instead, it appears to be part of a larger structure composed of randomly distributed field stars. 

The absence of clustering effects and the uniformity across all the regions as mentioned above, suggest that possibly the stars in the Eos cloud and its surroundings have been dynamically dispersed over time. Possible mechanisms for this dispersion include Galactic tidal interactions, past dynamical mixing, or dissolution of an older stellar association \citep{bland2016galaxy}. Future studies incorporating full three-dimensional velocity data (e.g., using radial velocity measurements from Gaia DR4 or spectroscopic surveys) could provide deeper insights into the origins and dynamical evolution of these stars. Moreover, this supports the statement that the Eos Cloud is not a site of active star formation but rather contains a population of field stars similar to other regions at this Galactic latitude.
 
\subsection{Potential for future star formation}
\cite{EosCloudPaper} calculated the overall Jeans stability of the Eos cloud, finding it to be marginally stable. They also calculated a global net molecular hydrogen dissociation and formation rate, finding that the cloud is net dissociating. \cite{Karoly2025} also concluded that the magnetic field strength is competitive with turbulence and gravity for the cloud. Overall  these initial expectations are consistent with our conclusions drawn based on the Gaia populations towards the cloud. However, molecular clouds/star formation are hierarchical processes and so an analysis of the stability and dynamics, as well as the molecular formation and dissociation rate on a range of spatial scales \citep[e.g.][]{2025arXiv250116474K} is required to fully conclude on the future star forming potential, or lack thereof, of the cloud. In particular, the presence of a CO-bright component (MBM 40) could be evidence of the beginning of collapse of a subset of the cloud. However, our study is in agreement with previous studies of the MBM 40 sub-region of the larger Eos cloud, indicating it is not currently star forming \citep{1996ApJ...465..825M}, and \citet{Karoly2025} determined a magnetic field strength in MBM 40 which is also competitive with gravity and turbulence. Further detailed study of the gas conditions and kinematics are required to fully determine the future star formation potential of the cloud.

\section{Summary and Conclusions}

The recently discovered Eos cloud \citep{EosCloudPaper} is one of the closest molecular clouds to the Sun, with the high latitude portion of the cloud beginning at 94pc. Although previously most of the cloud was undetected because it is CO-dark, it is abundant in molecular hydrogen and expected to be about half the mass of the Taurus/Lupus star forming regions. In this paper, we use Gaia data to study the stellar population towards the full extent of the Eos cloud to determine whether it has previously or recently undergone any star formation episodes. We find that: \\

\noindent i) Comparison of the observed Gaia CMD with pre-main sequence model isochrones imply there are no significant populations less than tens of Myr old within 25 degrees of the centre of the Eos cloud, over a distance 70-150\,pc from Earth (note that the Eos cloud extends from $\sim94-136$\,pc). \\

\noindent ii) There is no evidence of kinematic or spatial stellar clustering within the Eos cloud. \\

\noindent iii) The stellar population in analogous search volumes to the left and right of the Eos cloud (negative and positive longitudes, respectively) are statistically similar. \\

We therefore determine that the Eos cloud has not recently undergone any star formation episodes, consistent with the conclusions drawn from the magnetic field structure \citep{Karoly2025}.  Future work on the structure and dynamics of the cloud and the molecular formation/dissociation rate will help to reveal whether it may eventually be star forming. 

\section*{Acknowledgements}
We thank the referee for their timely and positive review of our manuscript.

TJH acknowledges UKRI guaranteed funding for a Horizon Europe ERC consolidator grant (EP/Y024710/1) and a Royal Society Dorothy Hodgkin Fellowship. 
EG gratefully acknowledges UK Research and Innovation (UKRI) guaranteed funding for a Horizon Europe ERC starting grant (EP/Z000890/1) and from the UK Science and Technology Facilities Council (STFC; project reference ST/W001047/1). TED. acknowledges support for this work provided by NASA through the NASA Hubble Fellowship Program grant No. HST-HF2-51529 awarded by the Space Telescope Science Institute, which is operated by the Association of Universities for Research in Astronomy, Inc., for NASA, under contract NAS 5-26555. K.P. is a Royal Society University Research Fellow, supported by Grant No. URF$\backslash$R1$\backslash$211322. J.K. is currently supported by the Royal Society under grant number RF\textbackslash ERE\textbackslash231132, as part of project URF\textbackslash R1\textbackslash211322.

\section*{Data Availability}

All Gaia data used in this paper is publicly available in the Gaia science archive\footnote{\url{https://gea.esac.esa.int/archive/}}. The FIMS/SPEAR data is publicly available in the MAST science archive \footnote{\url{https://archive.stsci.edu/missions-and-data/fims-spear}}.



\bibliographystyle{mnras}
\bibliography{example} 

\bsp	
\label{lastpage}
\end{document}